\documentclass[a4paper,12pt]{iopart}

\usepackage{iopams}
\usepackage{graphicx}

\begin{document}

\begin{flushright}
Imperial/TP/08/AC/01
\end{flushright}

\title{Non-Gaussianity from massless preheating}
\author{Alex Chambers and Arttu Rajantie}
\address{Department of Physics, Imperial College London, London SW7 2AZ, United Kingdom}
\ead{alex.chambers@imperial.ac.uk, a.rajantie@imperial.ac.uk}
\date{\today}

\bibliographystyle{h-physrev3}

\begin{abstract}
Preheating can convert superhorizon fluctuations of light scalar fields present at the end of inflation into observable density perturbations. We show in detail how lattice field theory simulations and the separate universes approximation can be used to calculate these perturbations and make predictions for the nonlinearity parameter $f_{\rm NL}$. We also present a simple approximation scheme that can reproduce these results analytically. Applying these methods to the massless preheating model, we determine the parameter values that are ruled out by too high levels of non-Gaussianity.
\end{abstract}
\pacs{98.80.Cq, 11.15.Kc}

\section{Introduction}
The non-Gaussianity of primordial density perturbations is a promising way to probe the very early universe. It can be used to determine the precise physics of inflation~\cite{Bartolo:2004if} and to distinguish between inflation and its alternatives \cite{Lehners:2007wc,Lehners:2008my}. Although observations are currently compatible with Gaussian perturbations~\cite{Komatsu:2008hk}, their sensitivity will improve appreciably over the next few years.

Whilst Gaussian perturbations are characterised completely by their two-point function, it is impossible to give a complete characterisation of non-Gaussian perturbations. Non-Gaussianity is usually parameterised phenomenologically by the nonlinearity parameter $f_{\rm NL}$, originally defined \cite{Komatsu:2000vy} for a specific `local' type of non-Gaussianity by
\begin{equation}
\label{equ:fNLorig}
 \zeta=\zeta_0-\frac{3}{5}f_{\rm
NL}\left(\zeta_0^2-\langle\zeta_0^2\rangle\right),
\end{equation}
where $\zeta$ is the curvature perturbation and $\zeta_0$ is a Gaussian random field. The definition of $f_{\rm NL}$ has since been extended to cover arbitrary non-Gaussian fields \cite{Maldacena:2002vr,Boubekeur:2005fj}, including cases in which it is scale dependent.

Current observational limits from the WMAP 5-year data show that the perturbations are close to Gaussian with no departure from Gaussianity \cite{Komatsu:2008hk}: $-9 <f_{\rm NL}^{\rm local} < 111$,  $-151 <f_{\rm NL}^{\rm equil.} < 253$ at 95\% confidence limit. However, a recent work \cite{Yadav:2007yy} using the WMAP three-year data found a significant non-Gaussianity: $26.9 <f_{\rm NL}^{\rm local} < 146.7$ at 95\% confidence limit. Measurements of large scale structure give $-29 <f_{\rm NL}^{\rm local} < 69$ at 95\% confidence limit \cite{Slosar:2008hx}. These ranges will be significantly reduced in the next few years with further WMAP data and data from the Planck satellite.

To make use of a future observation of non-Gaussianity, reliable predictions must be made by each inflationary model. 
Single field, slow-roll inflation predicts near-Gaussian primordial density perturbations, with $f_{\rm NL}$ of the order of the slow-roll parameters $\epsilon$ and $\eta$ \cite{Bartolo:2004if}. The perturbations generated during slow roll in multi-field models can be more non-Gaussian~\cite{Rigopoulos:2005us,Bernardeau:2002jy,Lyth:2005fi,Allen:2005ye}. 

Even larger values are possible in models with non-equilibrium dynamics during reheating~\cite{Kofman:1996mv,Dolgov:1982th}, the process which transfers energy from the inflaton field to matter and radiation after inflation has ended. The simplest example of this is the curvaton model~\cite{Linde:1996gt,Lyth:2001nq,Moroi:2001ct,Lyth:2002my}, in which curvaton particles dominate the energy density of the universe and therefore influence the expansion of the universe until they decay. 

In many models reheating starts with a brief period of non-equilibrium dynamics know as preheating~\cite{Traschen:1990sw,Kofman:1994rk}, which results in the rapid production of particles and very large occupation numbers. There have been many attempts to calculate the perturbations generated using linearised or semilinearised methods in, for example, the hybrid inflation model \cite{Barnaby:2006cq,Barnaby:2006km} and for chaotic inflation \cite{Bassett:1999cg,Finelli:2000ya,Enqvist:2004ey,Jokinen:2005by} and the separate universe approximation~\cite{Tanaka:2003cka,Nambu:2005qh,Suyama:2006rk}.

In a recent paper~\cite{Chambers:2007se} we developed a fully nonlinear calculation using lattice field theory simulations and the separate universe approximation. We showed that in the massless preheating model~\cite{Greene:1997fu} parametric resonance can produce very high levels of non-Gaussianity, ruling out parameter values. Conversely, other parameter values led to strong resonance but negligible non-Gaussianity.

In this paper we present a more comprehensive study of this model and full details of the simulations.

\section{Massless Preheating}
We study a simple variant of chaotic inflation known as massless preheating, 
the dynamics of which have been thoroughly studied previously \cite{Bassett:1999cg,Finelli:2000ya,Greene:1997fu}. The model consists of an inflaton field $\phi$ coupled to a massless scalar field $\chi$, with the potential
\begin{equation}
V(\phi,\chi)=\frac{1}{4}\lambda\phi^4+\frac{1}{2}g^2\phi^2\chi^2.
\end{equation}

During inflation $\chi$ is approximately zero and the model behaves the same way as the standard single field $\phi^4$ chaotic inflation model. 
We assume that the dominant contribution to the density perturbations is generated in the usual way by the inflaton field, so that the constraints arising from the linear perturbations are the same as for chaotic inflation. This fixes $\lambda$ to be $7\times 10^{-14}$ \cite{liddlelyth}, which we will use throughout this paper. This model is actually only marginally compatible with current observations~\cite{Komatsu:2008hk}, but we still choose to use it because of its convenience: The field dynamics are conformal \cite{Greene:1997fu}, which means that the relevant physics is at roughly the same comoving scale throughout our simulations. As an example, in the case of the potential $V(\phi,\chi)=\frac{1}{2}m\phi^2+\frac{1}{2}g^2\phi^2\chi^2$ the inverse mass $1/m$ sets a fixed (not comoving) physical length scale, which has to fit inside the simulation box at all times, requiring much larger lattices than those used here. Also, an analysis of the $\lambda \phi^4$ model using second order perturbation theory \cite{Jokinen:2005by} found $f_{\rm NL}\gtrsim\mathcal{O}(1000)$ in the parameter range $1<g^2/\lambda<3$.

Unless the coupling ratio $g^2/\lambda$ is large, the masses of the two fields, $m_\phi=\sqrt{3\lambda}\phi$ and $m_\chi=g\phi$, are comparable during inflation. Furthermore, $\chi$ is light relative to the Hubble rate $H$ except near the end of inflation. Quantum fluctuations of the $\chi$ field are amplified and stretched by inflation in the same way as those of the inflaton field. At the end of inflation, the $\chi$ field will therefore have an approximately scale invariant spectrum of perturbations (see \ref{app:spectra}).

The slow roll conditions fail and inflation ends when
 $\phi\approx 2\sqrt{3} M_{\rm Pl}$, where $M_{\rm Pl}=(8\pi G)^{-1/2}$ is the reduced Planck mass. After this, the inflaton fields starts to oscillate around zero with an amplitude decreasing from this value. The inflaton is massless leading the expansion of the universe to be approximately similar to radiation domination, $a(t)\propto t^{1/2}$. During these oscillations, the $\chi$ field resonates with the inflaton field $\phi$ transferring energy away from it, until the amplitude of $\chi$ becomes so high that the dynamics becomes nonlinear~\cite{Traschen:1990sw,Kofman:1994rk}. This process washes out the perturbations of $\chi$ produced during inflation and therefore no isocurvature modes survive. However, the perturbations affect the time the resonance lasts and, consequently, the amount of expansion during this period of preheating. This means that the perturbations of $\chi$ will leave an imprint in the curvature perturbation, and this contribution is generally non-Gaussian.

\section{Separate Universe Approximation}
\label{sec:sepuni}

Our approach~\cite{Chambers:2007se} combines classical lattice field theory 
methods~\cite{Khlebnikov:1996mc,Prokopec:1996rr}
with the widely used separate
universe approximation \cite{Lyth:2005fi,Starobinsky:1986fxa,Salopek:1990jq,Sasaki:1995aw,Lyth:2004gb}. It states that points in space separated by more than a Hubble distance cannot interact and will therefore evolve independently of each other. As long as each Hubble volume is approximately isotropic and homogeneous, one can approximate them by separate Friedmann-Robertson-Walker (FRW) universes. Gravitational effects are therefore described solely by Friedmann equation, which is nonlinear but does not take into account gradients and is therefore only valid at long distances.

In earlier applications of the this approximation ~\cite{Tanaka:2003cka,Nambu:2005qh,Suyama:2006rk} each separate universe was treated as if it was point-like. This means that not only the metric but also the fields were assumed to be homogeneous inside each separate universe. As we showed~\cite{Chambers:2007se}, this assumption is generally not valid during preheating. Instead, we allow the fields $\phi$ and $\chi$ to be inhomogeneous on small scales but we will assume that the metric has the usual homogeneous and isotropic FRW form. This approximation should be reasonably good as long as the size of our separate universes is not much larger than the horizon size, but ultimately it should be improved by including short-distance metric perturbations in the calculation. It is convenient to choose the separate universes to have a fixed comoving size $L$. For the FRW metric to be a good approximation for the whole lattice, this should be less than the comoving horizon size, $L<1/aH$, throughout the whole simulation.

The curvature perturbation $\zeta$ is given by the perturbation of the logarithm of the scale factor $a$ on a constant energy density hypersurface~\cite{Salopek:1990jq}
\footnote{In the literature it is more common to use lower case delta for this purpose, i.e., $\delta N$. However, we reserve $\delta$ for small-scale perturbations within one single separate universe and use $\Delta$ for large-scale perturbations between separate universes.}
\begin{equation}
\label{equ:deltan}
\zeta=\Delta N\equiv \Delta \ln a|_H,
\end{equation}
which one can find by solving the Friedmann equation independently for each separate universe. If different separate universes have different initial conditions, they evolve differently, and this will create curvature perturbations.

In this paper our emphasis is on non-Gaussianity produced at the end of inflation, and therefore we calculate the perturbations during slow roll inflation at linearized level. We switch to the separate universe picture when $\phi=\phi_{\rm ini}=5M_{\rm Pl}$, which is a few $e$-foldings before the end of inflation. Using a different value for $\phi_{\rm ini}$ should not change our overall results as long as it is in the slow-roll regime. The initial conditions for our separate universe calculation are therefore provided by the usual Gaussian field perturbations produced during inflation. 
In a given separate universe, we can write the initial field configurations as
\begin{eqnarray}
\phi(x)&=&\phi_{\rm ini}+\delta\phi(x),\nonumber\\
\chi(x)&=&\chi_{\rm ini}+\delta\chi(x),
\end{eqnarray}
where the mean fields $\phi_{\rm ini}$, $\chi_{\rm ini}$ are homogeneous over the each individual separate universe. They are determined by field perturbations with wavelength longer than $L$. We assume they vary from one separate universe to another with a Gaussian probability distribution determined by the power spectra of the fields. The fluctuations
$\delta\phi$ and $\delta\chi$ correspond to field perturbations with wavelength shorter than $L$, and therefore they are inhomogeneous within a single separate universe. They are represented by Gaussian random fields with zero mean and we assume they have the same statistical properties in each separate universe.

As $\phi$ and $\chi$ are the only light fields present at the end of inflation, all that needs to be considered for the calculation of the curvature perturbation $\zeta$ is the dependence of $N$ on their initial values. The inhomogeneous modes $\delta\phi(x)$ and $\delta\chi(x)$ do not give a direct contribution because they have the same statistics in each separate universe. Therefore, the curvature perturbation is determined by the initial mean values $\phi_{\rm ini}$ and $\chi_{\rm ini}$. What we therefore need to find is how $N$ depends on these two numbers $\phi_{\rm ini}$ and $\chi_{\rm ini}$.

Our model is symmetric under $\chi\rightarrow -\chi$, and $N$ will also possess this symmetry. The Taylor expansion of (\ref{equ:deltan}) can therefore only have even powers of $\chi_{\rm ini}$:
\begin{equation}
\label{equ:zetadef}
 \zeta(\phi_{\rm ini},\chi_{\rm ini})=\Delta N(\phi_{\rm ini},0)
+\frac{1}{2}\frac{\partial^2 N}{\partial\chi_{\rm ini}^2}\chi_{\rm ini}^2
	+O(\chi_{\rm ini}^4),
\end{equation}
where the second derivative is calculated at $\chi_{\rm ini}=0$. The first term $N(\phi_{\rm ini},0)$ is independent of $\chi_{\rm ini}$ and is therefore exactly the same as in the single-field $\phi^4$ model where curvature perturbations are known to be highly Gaussian. We will therefore focus on the $\chi_{\rm ini}^2$ term, which gives a non-Gaussian contribution.
To measure the level of this non-Gaussianity, we need to find its coefficient
$\partial^2 N/\partial\chi_{\rm ini}^2,$ which we do by measuring
the dependence of $N\equiv\ln a$ on $\chi_{\rm ini}$.

The contribution by the $\chi$ field to the non-Gaussianity is not of the simple `local' type (\ref{equ:fNLorig}), but defining $f_{\rm NL}$ as a suitable ratio of the three-point and two-point correlation functions of $\zeta$, a formula for the effective value of $f_{\rm NL}$ can be derived \cite{Boubekeur:2005fj,Lyth:2006gd,Lyth:2007jh}. Following Boubekeur and Lyth~\cite{Boubekeur:2005fj} the bispectrum can be defined as \cite{Komatsu:2001rj},
\begin{equation}
  B_\zeta \left( k_1,k_2,k_3\right) =-\frac{6}{5}f_{\rm NL}\left[P_\zeta\left( k_1\right)P_\zeta\left( k_2\right)+\rm{cyclic}\right].
\end{equation}
We assume that the contribution from the inflaton is practically Gaussian and dominates the power spectrum:
\begin{eqnarray}
\label{eqn:assumpions1}
 P_\zeta\left( k\right) =P_{\zeta_\phi}+P_{\zeta_\chi}\simeq P_{\zeta_\phi},\\
\label{eqn:assumpions2}
 B_\zeta\left( k_1,k_2,k_3\right)\simeq B_{\zeta_\chi}\left( k_1,k_2,k_3\right).
\end{eqnarray} 
It was shown \cite{Boubekeur:2005fj} that if the average of $\chi$ over our currently observable universe is negligible and perturbations of both fields are scale invariant, the non-Gaussianity parameter is
\begin{equation}
\label{equ:fNLcalc}
 f_{\rm NL}\approx-\frac{5}{48}\left(\frac{\partial^2N}{\partial\chi_{\rm ini}^2}\right)^3
 \frac{{\cal P}_\chi^3}{{\cal P}_\zeta^2}
		\ln\frac{k}{a_0H_0}.
\end{equation}
For $\phi^4$ inflation we would use in this approximation ${\cal P}_\chi=(H^2/4\pi^2)$ and ${\cal P}_\zeta=(V/24\pi^2\epsilon M_{\rm Pl}^4)$ where $\epsilon$ is the slow roll parameter $\epsilon=8M_{\rm Pl}^2/\phi^2$. This leads to
\begin{equation}
\label{eqn:blfnl}
f_{\rm NL}\approx-\frac{5}{9\pi^2}\left(\frac{\partial^2N}{\partial\chi_{\rm ini}^2}\right)^3\lambda
M_{\rm Pl}^6\ln\frac{k}{a_0H_0}.
\end{equation}
The logarithm reflects an infrared divergence, which is cut off by the length scale at which the averages in (\ref{equ:fNLcalc}) are computed, i.e., the maximum observable scale. It is a result of the assumption that the $\chi$ field has a flat power spectrum.

However, the perturbations are not exactly scale invariant, and this changes the result.
The relevant power spectra are those at the beginning of our simulations, a few $e$-foldings before the end of inflation. It is shown in~\ref{app:spectra} that
\begin{equation}
\label{eqn:blfnltilt1}
\fl
f_{\rm NL}\approx
\left\{\begin{array}{ll}
-\left(\frac{\partial^2N}{\partial\chi_{\rm ini}^2}\right)^3\lambda
M_{\rm Pl}^6\left(\frac{N_k}{N_{\rm sim}}\right)^{3(2-g^2/\lambda)}, & 
\mbox{if } g^2/\lambda<(3/2)N_{\rm sim},\\
-\left(\frac{\partial^2N}{\partial\chi_{\rm ini}^2}\right)^3\lambda
M_{\rm Pl}^6
\left(\frac{6e^2N_k}{g^2/\lambda}\right)^{-3g^2/\lambda}
\left(\frac{N_k}{N_{\rm sim}}\right)^6 e^{9N_{\rm sim}},
& \mbox{if } g^2/\lambda>(3/2)N_{\rm sim}
\end{array}\right.,
\end{equation}
where $N_k\approx 60$ is the number of $e$-foldings before the end of inflation when the largest scales we observe today left the horizon, and $N_{\rm sim}$ is the number of $e$-foldings before the end of inflation when we begin our simulations. For our choice of $\phi_{\rm ini}$ this is $N_{\rm sim}=\frac{25}{8}$. Note that we drop the logarithms and other factors of order 1.

In (\ref{equ:fNLcalc})--(\ref{eqn:blfnltilt1})  we assumed that the probability distribution of $\chi_{\rm ini}$ has zero mean, so that $\chi_{\rm ini}$ is zero on average. However, we can only measure density perturbations in our currently observable universe, and perturbations are therefore measured relative to the average value $\overline{\chi_{\rm ini}}$ over this volume.
In order for (\ref{eqn:blfnltilt1}) to be valid, 
this average $\overline{\chi_{\rm ini}}$ has to be small enough.
To account for this, we write
$\chi_{\rm ini}=\overline{\chi_{\rm ini}}+\Delta\chi_{\rm ini}$. Then (\ref{equ:zetadef}) becomes
\begin{equation}
\label{eqn:zeta2}
\zeta=\zeta_0+c\left(\overline{\chi_{\rm ini}}+\Delta\chi_{\rm ini}\right)^2,
\end{equation} 
where
\begin{equation}
c=\frac{1}{2}\frac{\partial^2N}{\partial\chi_{\rm ini}^2}.
\end{equation}
Expanding (\ref{eqn:zeta2}) and dropping the constant term gives,
\begin{equation}
\label{eqn:zeta3}
\zeta=2c\overline{\chi_{\rm ini}}\Delta\chi_{\rm ini}+c\Delta\chi_{\rm ini}^2.
\end{equation}
\begin{figure}
 \centering
 \includegraphics*[width=13cm]{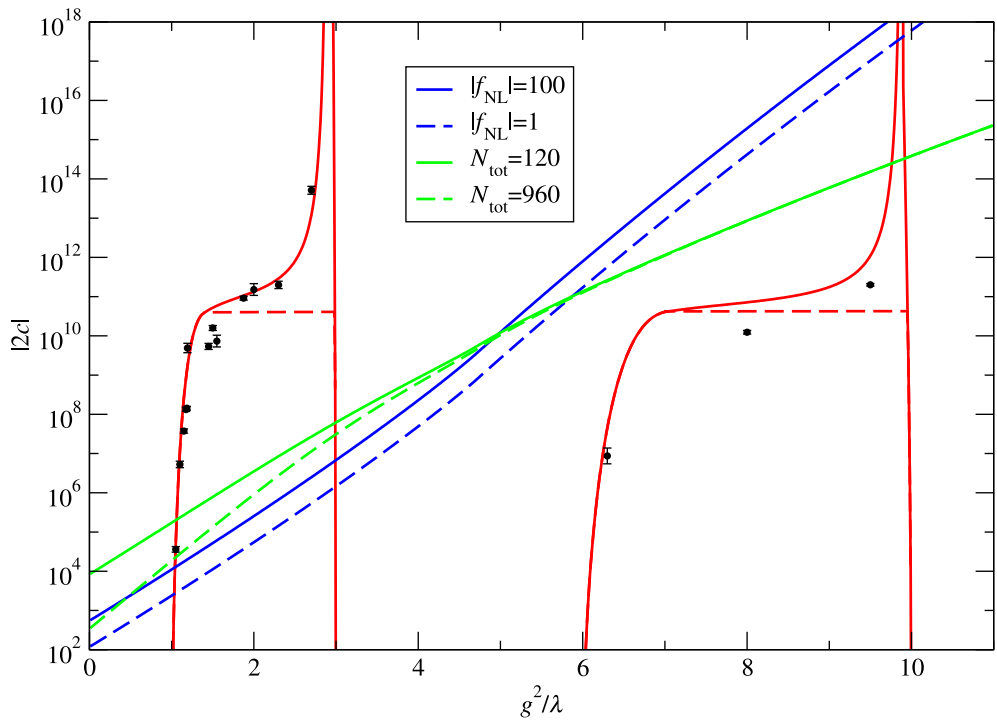}
 \caption{The black points are simulation data and the solid red (medium grey) line is the order of magnitude estimate given by (\ref{eqn:analyicfin}) for the box size of our simulations. The analytic prediction for an infinite box (for which the separate universes approximation is, of course, not valid) is given by the dashed red line. Under the assumptions (\ref{eqn:assumpions1}) and (\ref{eqn:assumpions2}) points above the blue (or dark grey) $f_{\rm NL}$ limits (derived from (\ref{eqn:blfnltilt1})) are excluded by observations. However, for points above the green (or light grey) lines (derived from (\ref{equ:cconstraint})) the same assumptions break down as preheating makes the dominant contribution to the power spectrum. This in itself is an interesting and non-trivial possibility. Note that at small $g^2/\lambda$ this limit depends on the total amount of inflation.
 }
 \label{fig:redline}
\end{figure}
The first term in (\ref{eqn:zeta3}) 
gives a Gaussian contribution to the curvature perturbation. 
If this has a smaller amplitude than the contribution from the inflaton field, 
\begin{equation}
\label{equ:cconstraint}
2c\overline{\chi_{\rm ini}}\Delta\chi_{\rm ini}\lesssim10^{-5},
\end{equation}
where $\Delta\chi_{\rm ini}$ refers to the typical (root mean squared) perturbation at the horizon scale,
it does not affect observations. Therefore the average $\overline{\chi_{\rm ini}}$ can be safely ignored. 

On the other hand, if the constraint (\ref{equ:cconstraint}) is not satisfied, preheating is the dominant source of curvature perturbations. This means that we would have to use a smaller coupling $\lambda$ to obtain the observed amplitude, and other observables such as the spectral index would also be different from the usual predictions.

The typical values of $\overline{\chi_{\rm ini}}$ and $\Delta\chi_{\rm ini}$ are given by the variances
\begin{equation}
\label{equ:varchibar}
 \left\langle \overline{\chi_{\rm ini}}^2\right\rangle =\int^{a_0H_0}_0\mathcal P_\chi\left(k\right)\frac{dk}{k},
\end{equation}
and
\begin{equation}
\label{equ:vardelchi}
 \left\langle \Delta{\chi_{\rm ini}}^2\right\rangle =\int^{H}_{a_0H_0}\mathcal P_\chi\left(k\right)\frac{dk}{k},
\end{equation}
where $H_0$ is the Hubble parameter today, $a_0$ is the scale factor today and $H$ is the Hubble parameter at the start of our simulations. For $g^2/\lambda<2$ the spectrum, $\mathcal P_\chi\left(k\right)$, is red tilted (there is more power at larger scales) and the integral is infrared divergent.
As we show in \ref{app:spectra}, this means that the constraint depends on the total amount of inflation $N_{\rm tot}$. For $g^2/\lambda\rightarrow 0$, we find
\begin{equation}
\label{equ:smallglimit}
\left|\frac{\partial^2N}{\partial\chi_{\rm ini}^2}\right|\lesssim \frac{9\pi^2}{4\lambda}(N_{\rm tot}N_0)^{-3/2}\times 10^{-5}
\approx 3000\times (N_{\rm tot}/100)^{-3/2},
\end{equation}
where $N_0\approx 60$ is the number of e-foldings of inflation after largest currently observables scales left the horizon.
For $g^2/\lambda>2$ the spectrum is blue tilted (there is less power at larger scales) and the integral converges. Because of this, the constraint on $c$ becomes rapidly much weaker, as shown in figure~\ref{fig:redline}. See \ref{app:spectra} for further discussion of this issue.

\section{Analytic Approximation}
\label{sec:ana}

In this section we derive an analytic approximation of the second derivative $\partial^2N/\partial\chi_{\rm ini}^2$, and therefore $f_{\rm NL}$, by linearising the field dynamics of preheating. This will give an order of magnitude estimate to compare with the field theory simulations described in section \ref{sect:sims}.

 \begin{figure}
  \centering
  \includegraphics[width=10cm]{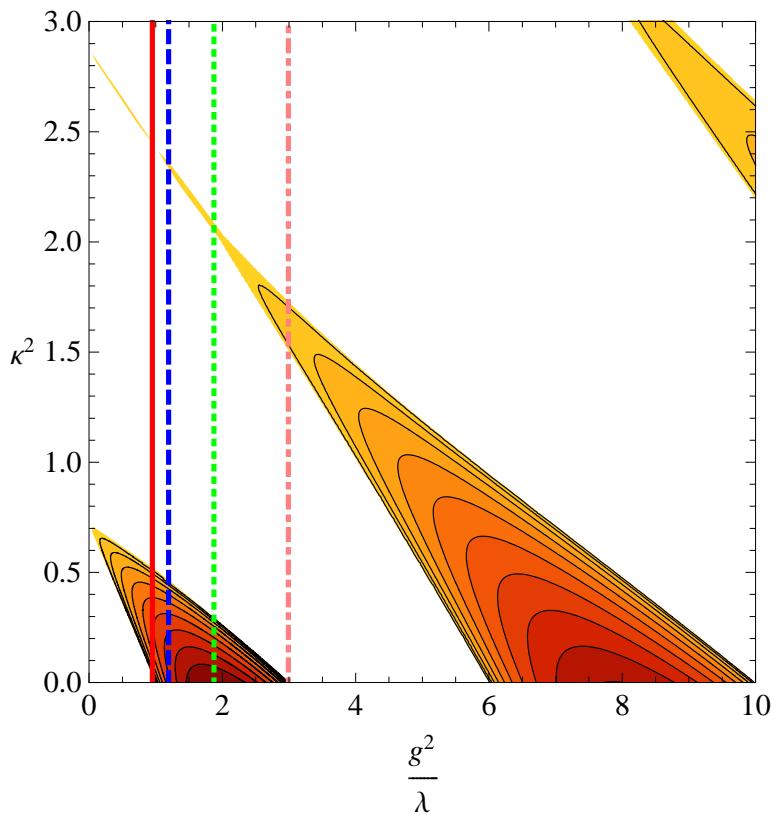}
  \caption{Resonance structure of the massless preheating model. The contours show the Floquet index, $\mu$, from the solution of (\ref{equ:lame}). The peak of the first band is at
 $g^2/\lambda=1.875$. The four marked sections are shown in figure \ref{fig:floquet2_log}}
  \label{fig:floquet1}
 \end{figure}

The dynamics of the fields are described by a system of two coupled ODEs:
\begin{eqnarray}
\label{phi_eq_motion}
\ddot\phi+3H\dot\phi-\frac{1}{a^2}
\vec\nabla^2\phi+\lambda\phi^3+g^2\phi\chi^2&=&0,\\
\label{chi_eq_motion}
\ddot\chi+3H\dot\chi-\frac{1}{a^2}\vec\nabla^2\chi+g^2\phi^2\chi&=&0.
\end{eqnarray}
The details of the field dynamics are given in detail in \cite{Greene:1997fu}, but we will repeat some of the more important points here. At the end of inflation the inflaton field, $\phi$, reaches the bottom of its potential and starts to oscillate with decreasing amplitude. During and immediately after inflation the $\chi$ field is approximately zero and we can ignore the terms involving $\chi$ in (\ref{phi_eq_motion}), along with the gradient term as the field is approximately homogeneous. In terms of rescaled field $\tilde\phi=a\phi$ and rescaled conformal time $\tilde\tau$ defined by $d\tilde\tau=a^{-1}\lambda^{1/2}A_{\phi}dt$, the oscillations are approximately described by the Jacobi cosine function\footnote{We follow the usual convention in cosmology and some mathematical works \cite{htbook} and use $1/\sqrt{2}$ in the second argument of the Jacobi cosine function. Note that the same function is often given as ${\rm cn}(\tilde\tau\vert\frac{1}{2})$ \cite{abramsteg}.} \cite{htbook},
\begin{equation}
\label{eqwithsqroot}
 \tilde\phi(\tilde\tau)=A_{\phi}{\rm cn}(\tilde\tau,1/\sqrt{2}),
\end{equation}
where $A_{\phi}$ is the constant amplitude of the $\tilde\phi$ oscillations. This is set to $A_\phi =2\sqrt{3}M_{\rm Pl}$, the value of $\phi$ when slow roll is violated. The oscillations in the $\phi$ field give rise to an oscillatory mass term for the $\chi$ field. At linear level, a Fourier mode of the rescaled field $\tilde\chi=a\chi$
with wave number $k$ satisfies the Lam\'e equation,
\begin{equation}
\label{equ:lame}
\tilde\chi_k''+\left[
			\kappa^2+\frac{g^2}{\lambda}{\rm
cn}^2(\tilde\tau,1/\sqrt{2})\right]
			\tilde\chi_k=0, 
\quad
\kappa^2=\frac{k^2}{\lambda A_{\phi}^2}.
\end{equation}

\begin{figure}
 \centering
 \includegraphics*[width=12cm]{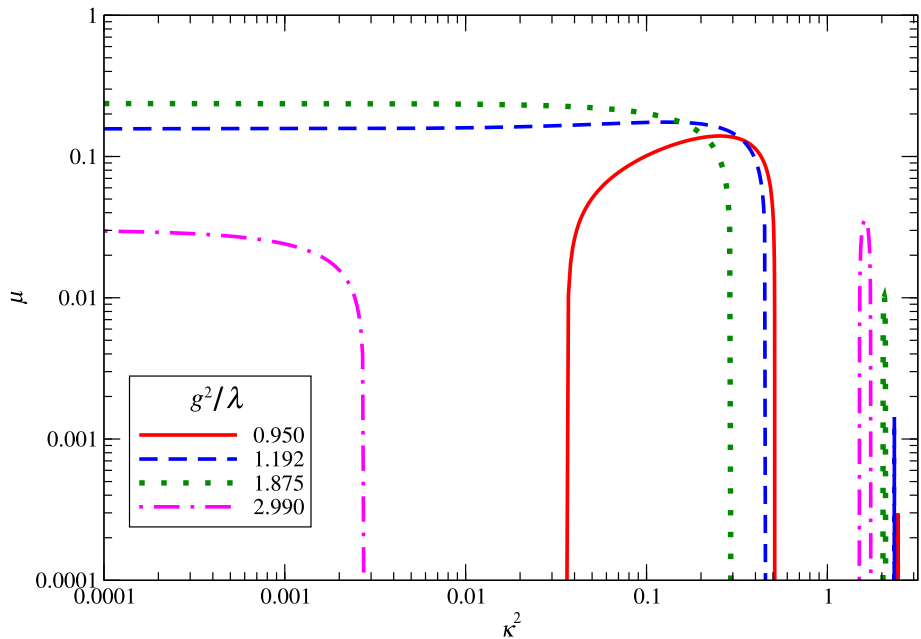}
 \caption{Cross-sections of the resonance structure shown in figure \ref{fig:floquet1} at constant values of $g^2/\lambda$.}
 \label{fig:floquet2_log}
\end{figure}

In the space of the two constant parameters $\kappa^2$ and $g^2/\lambda$, (\ref{equ:lame}) has resonance bands in which the solution grows
exponentially, 
\begin{equation}
\label{equ:Floquetgrowth}
\tilde\chi_k(\tilde\tau)=e^{\mu(\kappa,g^2/\lambda)\tilde\tau}f(\tilde\tau),
\end{equation} 
where $\mu(\kappa,g^2/\lambda)$ is known as
the Floquet index and $f(\tilde\tau)$ is a periodic function. 
This means that energy is transferred rapidly from the inflaton, $\phi$, to the
$\chi$ field. The Floquet index for a particular set of $\kappa^2$ and
$g^2/\lambda$ can be calculated by numerically solving (\ref{equ:lame})
over one period of oscillation and finding the eigenvalues of the matrix that relates the final and initial values.

As figure \ref{fig:floquet1} shows, the resonance bands stretch through the $(g^2/\lambda,\kappa^2)$ plane diagonally so that for any value of $g^2/\lambda$ there are some resonant modes.  Figure \ref{fig:floquet2_log} shows how the Floquet index depends on $\kappa$ at some chosen values of $g^2/\lambda$. When $g^2\ll\lambda$, the bands are narrow and the resonance is weak. Effective preheating therefore requires $g^2/\lambda \gtrsim 1$.

It is important for our purposes here to note when the modes of the $\chi$ field grow on large scales. This is when modes with $\kappa^2=0$ fall in the resonance bands in figure \ref{fig:floquet1}. The edges of these resonance bands fall at $g^2/\lambda=\frac{1}{2} n\left(n+1\right)$ \cite{Greene:1997fu}; the first band is $1<g^2/\lambda<3$, the second is $6<g^2/\lambda<10$ and so on.

To estimate the curvature perturbation produced by the resonance, we describe the resonance using the linear approximation. When the amplitude of $\chi$ has grown so much that this approximation is no longer valid, we assume that the resonance stops and the fields equilibrate instantaneously.

In the first instance we assume that the universe expands with the radiation-dominated equation of state, so that $a\propto t^{1/2}$. In terms of rescaled conformal time $\tilde\tau$, the scale factor is
\begin{equation}
a(\tilde\tau)=1+\frac{A_{\phi}}{\sqrt{12}M_{\rm Pl}}\tilde\tau.
\end{equation}
In the following we will use the scale factor $a$ as the time coordinate rather than $\tilde\tau$.

In the linear approximation field perturbations remain Gaussian.
This means that the modes are independent and the zero mode and inhomogeneous modes can be separated,
\begin{equation}
\label{eqn:splitmodes}
 \left\langle \tilde\chi^2\right\rangle \left( a\right)
=\overline{\tilde\chi}^2\left( a\right)+\left\langle
\delta\tilde\chi^2\right\rangle \left( a\right),
\end{equation} 
where $\tilde\chi=a\chi$. 

Following (\ref{equ:Floquetgrowth}), we approximate the evolution of mode with comoving wavenumber $k$ by
$\tilde\chi_k(a)\propto e^{\tilde\mu_k a},$
 where $\tilde\mu=\sqrt{12}\mu M_{\rm Pl}/A_{\phi}$ is a rescaled Floquet index.
Consequently, the first term in (\ref{eqn:splitmodes}) grows as
\begin{equation}
\label{eqn:approxhom}
\overline{\tilde\chi}^2\left( a\right)=\overline{\tilde\chi}^2\left( 1\right)e^{2\tilde\mu_0\left( a-1\right)}=\tilde\chi^2_{\rm ini}e^{2\tilde\mu_0\left( a-1\right)}.
\end{equation}
Note that for our choice of $a_{\rm ini}=1$, $\tilde\chi_{\rm ini}=\chi_{\rm ini}$ so we drop the tilde. We approximate the evolution of the second term by,
\begin{equation}
\label{eqn:approxinhom}
 \left\langle\delta\tilde\chi^2\right\rangle \left( a\right)
=\int \frac{d^3k}{\left( 2\pi\right)^3}\frac{1}{2k}
\left(e^{2\tilde\mu_{k}\left( a-1\right)}-1\right)
\sim m^2e^{2\tilde\mu_{\rm max}\left( a-1\right)},
\end{equation}
where $m^2\equiv \lambda A_{\phi}^2$  
 and $\tilde\mu_{\rm max}$ is the Floquet index of the mode with the largest Floquet index for a choice of $g^2/\lambda$. 
More precisely, the non-exponential prefactor is determined by the shape of the resonance band and it therefore depends on the coupling ratio $g^2/\lambda$. It would even be reasonably straightforward to compute it numerically by evaluating the integral. However, since the only relevant dimensionful scale is $m$, this serves as a sufficient order-of-magnitude estimate.

\begin{figure}
 \centering
 \includegraphics*[height=8cm]{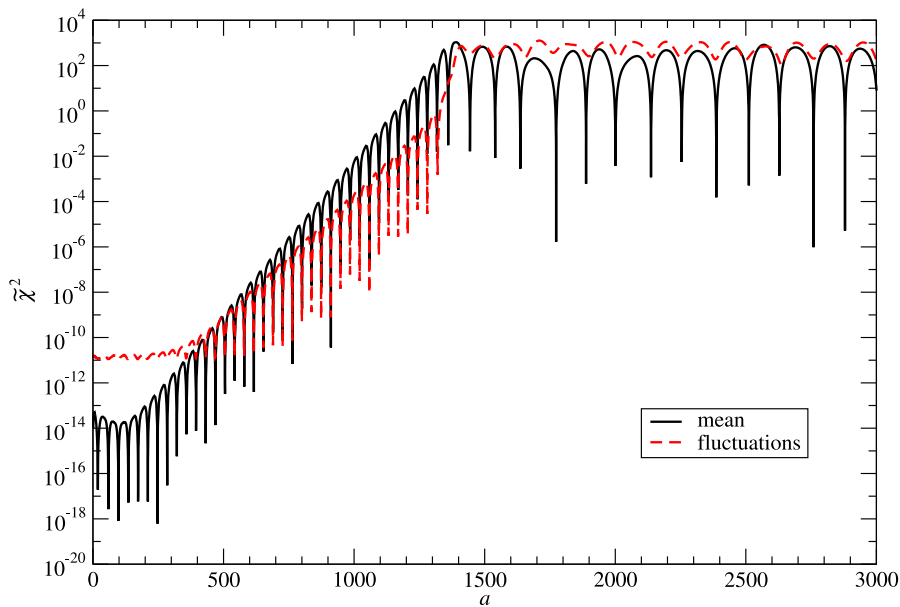}
 \caption{Evolution of the $\chi$ field during one simulation for $g^2/\lambda=2.7$.}
 \label{fig:modes}
\end{figure}

Also by inspection of (\ref{phi_eq_motion}) it can be seen that the system becomes nonlinear at scale factor $a_{\rm nl}$ when $g^2\left\langle \tilde\chi^2(a_{\rm nl})\right\rangle\simeq \lambda\tilde\phi^2\simeq m^2$ (see also figure \ref{fig:modes}). Substituting (\ref{eqn:approxhom}) and (\ref{eqn:approxinhom}) into (\ref{eqn:splitmodes}) gives,
\begin{eqnarray}
 m^2&=&g^2\left( \chi^2_{\rm ini}e^{2\tilde\mu_0\left( a_{\rm nl}-1\right)}+m^2e^{2\tilde\mu_{\rm max}\left( a_{\rm nl}-1\right)}\right)\nonumber\\
\label{eqn:an1}
&\simeq&g^2\left( \chi^2_{\rm ini}e^{2\tilde\mu_0a_{\rm nl}}+m^2e^{2\tilde\mu_{\rm max}a_{\rm nl}}\right).
\end{eqnarray}

The nonlinearity scale factor $a_{\rm nl}$ depends on $\chi_{\rm ini}$, and to determine it we Taylor expand it as
\begin{equation}
\label{equ:anlexp}
a_{\rm nl}(\chi_{\rm ini})=(1+c_{\rm nl}\chi_{\rm ini}^2)a_{\rm nl}(0)+O(\chi_{\rm ini}^4).
\end{equation}
We first solve the equation for $\chi_{\rm ini}=0$, and find the corresponding value $a_{\rm nl}(0)$
\begin{equation}
\label{eqn:calcazero}
 m^2 \simeq g^2m^2e^{2\tilde\mu_{\rm max}a_{\rm nl}(0)}\Rightarrow a_{\rm nl}(0) \simeq \frac{1}{\tilde\mu_{\rm max}}\ln\frac{1}{g}
\end{equation}
We then determine $c_{\rm nl}$ by substituting (\ref{equ:anlexp}) to (\ref{eqn:an1}),
\begin{equation}
 m^2\simeq g^2\left( \chi_{\rm ini}^2e^{2\tilde\mu_0a_{\rm nl}(0)}
 +m^2e^{2\tilde\mu_{\rm max}a_{\rm nl}(0)}e^{2\tilde\mu_{\rm max}a_{\rm nl}(0)c_{\rm nl}\chi_{\rm ini}^2}\right)
 +O(\chi_{\rm ini}^4).
\end{equation}
Rearranging this gives
\begin{equation}
c_{\rm nl}\simeq-\frac{1}{2\tilde\mu_{\rm max}a_{\rm nl}(0)m^2}e^{2\left( \tilde\mu_0-\tilde\mu_{\rm max}\right)a_{\rm nl}(0)},
\end{equation}
and substituting for $a_{\rm nl}(0)$ from (\ref{eqn:calcazero}) and $m^2=\lambda A_{\phi}^2$,
\begin{equation}
\label{eqn:cfin}
 c_{\rm nl}
 \simeq \frac{1}{2A_{\phi}^2\ln\frac{1}{g}}\frac{g^2}{\lambda}g^{-2\frac{\mu_0}{\mu_{\rm max}}}.
\end{equation}

Equations (\ref{equ:anlexp}) and (\ref{eqn:cfin}) tell us the value of the scale factor at the time when the dynamics becomes nonlinear. This, however, does not yet give the curvature perturbation. According to (\ref{equ:deltan}), the curvature perturbation is given by the scale factor at some fixed energy density or, equivalently, some fixed Hubble rate, which we denote by $H_*$. The result does not depend on the choice of $H_*$ as long as it is chosen to be after preheating has ended.

To calculate the curvature perturbation from (\ref{equ:anlexp}), we assume that once the dynamics has become nonlinear, the equation of state is exactly that of radiation, and therefore $H(a)=(a_{\rm nl}^2/a^2) H_{\rm nl}$, where $H_{\rm nl}$ is the Hubble rate at $a_{\rm nl}$.
Inverting this we find
$a(H_*)=\sqrt{H_{\rm nl}/H_*}a_{\rm nl}$.
This implies
\begin{equation}
\Delta\ln a(H_*)=\frac{1}{2}\Delta \ln H_{\rm nl}
+\Delta\ln a_{\rm nl}
= \left(1+\frac{1}{2}\frac{d \ln H}{d \ln a}\right) \Delta\ln a_{\rm nl}.
\end{equation}
The second derivative at constant $H$ in (\ref{equ:zetadef}) is therefore
\begin{equation}
\label{equ:anaresult}
\frac{\partial^2 N}{\partial\chi_{\rm ini}^2}=\left(1+\frac{1}{2}\frac{d \ln H}{d \ln a}\right)2c_{\rm nl}.
\end{equation}

If the equation of state during the resonance were exactly that of radiation, the two terms inside the brackets would cancel exactly and there would be no contribution to the curvature perturbation (\ref{equ:deltan}). However, the expansion rate depends on the phase of the oscillations, and in fact we have
\begin{equation}
\frac{d\ln H}{d\ln a}\approx -\frac{1}{2M_{\rm Pl}^2}\left(\frac{d\tilde\phi}{da}\right)^2=
-\frac{6}{A_{\phi}^2}\left(\frac{d\tilde\phi}{d\tilde\tau}\right)^2.
\end{equation}
Using (\ref{eqwithsqroot}), we find
\begin{equation}
\label{equ:osceos}
\frac{d\ln H}{d\ln a}\approx -6\left(\frac{d}{dt}{\rm cn}(\tilde\tau,1/\sqrt{2})\right)^2.
\end{equation}
This oscillates between $-3$ and $0$, and therefore
\begin{equation}
-\frac{1}{2}\le \left(1+\frac{1}{2}\frac{d \ln H}{d \ln a}\right) \le 1.
\end{equation}

This demonstrates that the two terms in (\ref{equ:anaresult}) do not generally cancel, and that in fact the conversion factor is or order one but can have either sign. Therefore we adopt as our analytic prediction
\begin{equation}
\label{eqn:analyicfin}
\frac{\partial^2 N}{\partial\chi_{\rm ini}^2}\approx \pm
2c_{\rm nl}
\approx \pm \frac{1}{A_{\phi}^2\ln\frac{1}{g}}\frac{g^2}{\lambda}g^{-2\frac{\mu_0}{\mu_{\rm max}}}.
\end{equation}
This is shown in figure~\ref{fig:redline} and its shape can be understood by comparing with the Floquet indices shown in table~\ref{tab:results}. The effect is only present when the zero mode is within a resonance band, i.e., $\mu>0$. Near the lower end of the resonance band, when $1<g^2/\lambda\lesssim 1.4$ (see the blue long-dashed line in figure~\ref{fig:floquet2_log}), the zero mode resonates but the maximum Floquet index is achieved at $\kappa>0$, and therefore $\mu_0/\mu_{\rm max}<1$.

Above this, for $1.4\lesssim g^2/\lambda\lesssim 2.986$ the resonance is at its strongest at $\kappa=0$. For the solid red line in figure~\ref{fig:redline} we have chosen to use as $\mu_{\rm max}$ the value for the longest wavelength available in our finite simulation box, to allow direct comparison with our numerical results. However, this makes the result dependent on the box size and therefore obviously unphysical. Taking the naive infinite-volume limit $L\rightarrow\infty$ would give us $\mu_0/\mu_{\rm max}=1$ (red dashed line),
but this would violate the constraint $L<H^{-1}$. Therefore, we have to accept that our approach is not reliable in this range of $g^2/\lambda$ and that the true value is probably somewhere between these two limits. Nevertheless, setting $\mu_0/\mu_{\rm max}=1$ gives a useful rule of thumb: $\partial^2N/\partial\chi_{\rm ini}^2\sim 1/\lambda$ when $g^2/\lambda$ is close to the top of a resonance band. Therefore, we can see that $\partial^2N/\partial\chi_{\rm ini}^2$ will be large because inflation constrains $\lambda$ to be small ($\sim 10^{-14}$).

Finally, in the range $g^2/\lambda\gtrsim 2.986$, the strongest resonance is found in the second resonance band (pink dot-dashed line in figure~\ref{fig:floquet2_log}), and therefore the prediction falls very steeply in this range.

\begin{figure}
 \centering
 \includegraphics*[height=8cm]{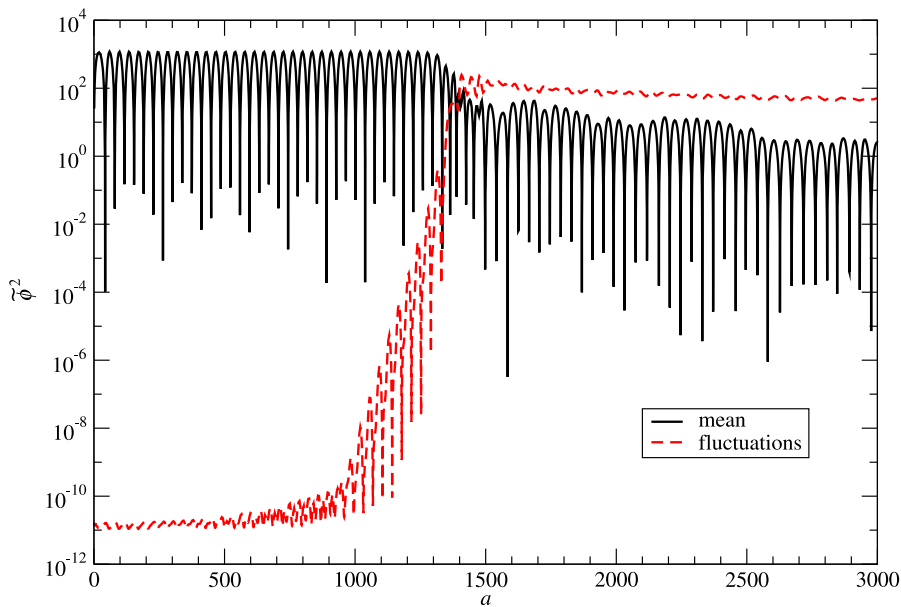}
 \caption{Evolution of the $\phi$ field during one simulation for $g^2/\lambda=2.7$.}
 \label{fig:phi}
\end{figure}

\section{Simulations}
\label{sect:sims}

In the simulations we employ three-dimensional classical field theory lattices. This is a standard method of solving such systems \cite{Khlebnikov:1996mc,Prokopec:1996rr,Felder:2000hq,Rajantie:2000fd,Copeland:2002ku,BasteroGil:2007mm},
but before \cite{Chambers:2007se}, it had not been used in the context of the separate universe approximation.
To determine the expansion of the universe,
we couple the equations of motion (\ref{phi_eq_motion}) with the Friedmann equation
\begin{equation}
\label{equ:inhomo3}
H^2=\frac{\overline\rho}{3M_{\rm Pl}^2},
\end{equation}
where the energy density is calculated as the average energy density in the simulation box,
\begin{equation}
\label{equ:densityint}
\overline\rho=\frac{1}{L^3}\int
d^3x\Bigl[\frac{1}{2}\dot\phi^2+\frac{1}{2}\dot\chi^2
+\frac{1}{2a^2}\left((\vec\nabla\phi)^2+(\vec\nabla\chi)^2\right)+V(\phi,
\chi)\Bigr].
\end{equation}
The coupled system of equations (\ref{phi_eq_motion}) and (\ref{equ:inhomo3}) are solved on a comoving lattice with periodic boundary conditions. We solved them in conformal time ($d\tau=a^{-1}dt$) using the second-order Runge-Kutta algorithm for the field equations coupled to an Euler method for the Friedmann equation. (Details of the algorithm used are presented in \ref{app:rk}). After each Runge-Kutta timestep for the fields the integral (\ref{equ:densityint}) is performed and the Euler timestep for $a$ is made. While it is possible to use a Runge-Kutta system to solve all the equations, we found the different order of the equations ((\ref{phi_eq_motion}) is second order and (\ref{equ:inhomo3}) is first order) leads to a cumulative error in $a$ and, therefore, numerical errors in the final results. This is not the case with the Runge-Kutta-Euler hybrid algorithm used here.
 
The separate universes approximation has been applied to massless preheating previously \cite{Tanaka:2003cka,Nambu:2005qh,Suyama:2006rk,Podolsky:2002qv}, but without including the field gradient terms in (\ref{phi_eq_motion}).
In these works the initial $\chi$ is varied and the log of the scale factor, $N$, at some later $H$ is found. These works found the function $N\left( \chi_{\rm ini}\right)\vert_H $ to be random, suggesting chaotic dynamics. However, the calculation does not include contributions from the inhomogeneous modes. It is the averaged dynamics which lead the function $N\left( \chi_{\rm ini}\right)\vert_H $ to depend on an average across all of the modes. As we shall see this addition of additional degrees of freedom leads to the chaotic dynamics being smoothed for sufficiently small values of $\chi_{\rm ini}$.

\begin{figure}
 \centering
  \includegraphics*[width=10cm]{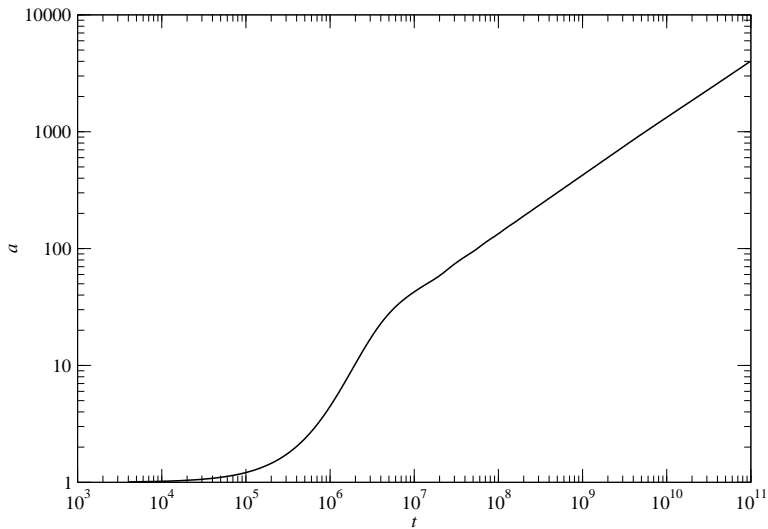}
 \caption{Evolution of the scale factor during one simulation. We begin the system a few ($\sim 3$) $e$-foldings before the end of inflation. During preheating the scale factor evolves approximately as in radiation domination.}
 \label{fig:scalefactor}
\end{figure}

In setting up the initial conditions, we treat the homogeneous and inhomogeneous modes differently. The lattice average of $\chi$ is set equal to $\chi_{\rm ini}$. For the inhomogeneous modes,
we follow the standard approach~\cite{Khlebnikov:1996mc,Prokopec:1996rr,Felder:2000hq,Rajantie:2000fd,Copeland:2002ku}. The $\chi$ field is given random initial conditions from a Gaussian distribution whose two-point functions are the same as those in the tree-level quantum vacuum state,
\begin{eqnarray}
\label{equ:quantumfluct}
\langle \chi_k\chi_q\rangle&=&(2\pi)^3\delta(k+q)\frac{1}{2\omega_k},\nonumber\\
\langle \dot\chi_k\dot\chi_q\rangle&=&(2\pi)^3\delta(k+q)\frac{\omega_k}{2},
\end{eqnarray}
where $\omega_k=\sqrt{k^2+m_\chi^2}=\sqrt{k^2+g^2\phi_{\rm ini}^2}.$
All other two-point correlators vanish. 
Early in the simulation (at the end of inflation and the beginning of preheating) the dynamics are linear, and the time evolution of this classical ensemble is identical to that in the quantum theory.
Later, when the field evolution becomes nonlinear the occupation numbers are so large that the classical theory is also a good approximation to the quantum field dynamics.

The inhomogeneous modes of $\phi$ are populated similarly to $\chi$. The initial conditions for homogeneous mode of the $\phi$ field is $\phi_{\rm ini}=5M_{\rm Pl}$. This is sufficient to drive $\sim 3$ $e$-foldings of inflation. The initial scale factor is $a=1$. The usual slow roll techniques (which are valid only during this early part of the simulation where $\phi\gtrsim 2\sqrt{3}M_{\rm Pl}$) are then employed to give the initial first derivative of $\phi$ with respect to conformal time $\tau$ to be $(d\phi/d\tau)_{\rm ini}=-\frac{\sqrt{4}}{3}\lambda \sqrt{M_{\rm Pl}}\phi_{\rm ini}$. From this point the Runge-Kutta-Euler algorithm described in \ref{app:rk} is used to solve (\ref{phi_eq_motion}) and  (\ref{equ:inhomo3}).

Figure \ref{fig:phi} shows the evolution of the mean of the inflaton field $\phi$ during one simulation time. The field begins above $M_{\rm Pl}$ which drives the initial inflation which can be seen in figure \ref{fig:scalefactor}. It rolls to the bottom of the potential and proceeds to oscillate about the minimum.
Example output for the $\chi$ field is shown in figure \ref{fig:modes}. 

We fixed the inflaton self coupling to $\lambda=7\times 10^{-14}$, which, assuming that inflation makes the dominant contribution to the power spectrum, gives the observed level of CMB fluctuations~\cite{liddlelyth}.  We used lattices of $32^3$ points with comoving spacing $\delta x=1.25\times 10^5$ and time step $\delta \tau=4\times 10^3$ in Planck units.
The lattice size is smaller than in most other studies of preheating because the calculation of curvature perturbation requires a large number of runs for each set of parameters, whereas in other, much less ambitious studies a few runs have been sufficient. It should also be noted that the constraint $L<1/aH$ limits the size of our lattice, and even with the current choice, the comoving horizon size is briefly somewhat smaller than $L$. Therefore, the number of points could only be increased by reducing $\delta x$, i.e., making the lattice finer. However, the relevant physical scales are longer than $\delta x$, and we see no indication that this would improve our results.

It should be noted that the gradient energy of the `quantum' fluctuations (\ref{equ:quantumfluct}) gives an ultraviolet vacuum energy contribution to the energy density. This can be estimated to be $\rho_{\rm UV}\approx \delta x^{-4}\sim 10^{-21}$, and we have to make sure that this is much less than the physical energy density stored in the inflaton potential $\rho_{\rm phys}\approx \frac{1}{4} \lambda \phi_{\rm ini}^4\sim 10^{-11}$. 

According to (\ref{equ:zetadef}), the contribution to the curvature perturbation from preheating depends on the second derivative  $\partial^2 N/\partial\chi_{\rm ini}^2$ at constant $H=H_*$, where $\chi_{\rm ini}$ is the value of the zero mode at the beginning of the simulation. 
If we calculate the second derivative at late enough times, when the system has reached a quasi-equilibrium state, the result should be independent of $H_*$ because the curvature perturbation is conserved.

\begin{figure}
 \centering
 \includegraphics*[width=10cm]{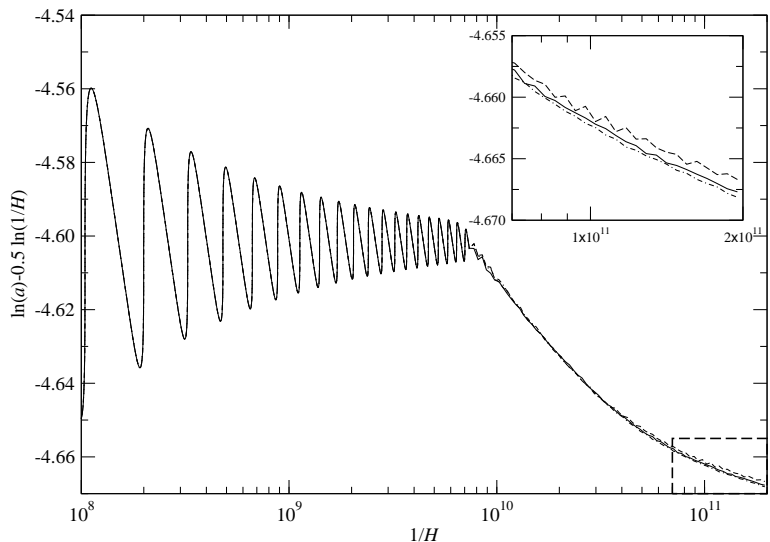}
 \caption{The evolution of the log of the scale factor, $\ln(a)$, at the end of a sample of simulations for $g^2/\lambda=1.875$, with radiation-domination-like expansion divided out. Each line averaged over 100 runs. The three curves are for different $\chi_{\rm ini}$.  Dashed: $\chi_{\rm ini}=0$. Dot-dashed: $\chi_{\rm ini}=1.0\times 10^{-7}$. Solid: $\chi_{\rm ini}=1.9\times 10^{-7}$. At late times these tend back towards radiation domination (a horizontal line in this plot). In the inset it can be seen that the lines a parallel at late times, resulting in the perturbations being locked in.}
 \label{fig:h}
\end{figure}

Our simulations give us the functions $a(\tau)$ and $H(\tau)$, from which we obtain $a(H)$. This is shown in figure \ref{fig:h}, where we have subtracted the underlying radiation-dominated evolution $a\propto H^{-1/2}$. 
Initially, we can see the effect of the coherent oscillations of the $\phi$ field, which correspond to (\ref{equ:osceos}). When the dynamics become nonlinear at $H\sim 10^{-10}M_{\rm Pl}$, these die away. However, as the inset shows, smaller oscillations due to non-equilibrium effects and statistical errors remain.

To find $\partial^2 N/\partial\chi_{\rm ini}^2$ we need the scale factor for different $\chi_{\rm ini}$ at some chosen value $H=H_*$, whereas the output from the code is at discrete values of $H$ which are different for each run. Therefore we have to interpolate the data from each run to $H=H_*$. To simultaneously remove the effect of the transient oscillations, we fit a power-law function to the data over a range of $1/H$ that is longer than the characteristic length of the transient oscillations, and use that to determine $N=\ln a(H_*)$ for each $\chi_{\rm ini}$.

\begin{figure}
 \centering
 \includegraphics*[width=14cm]{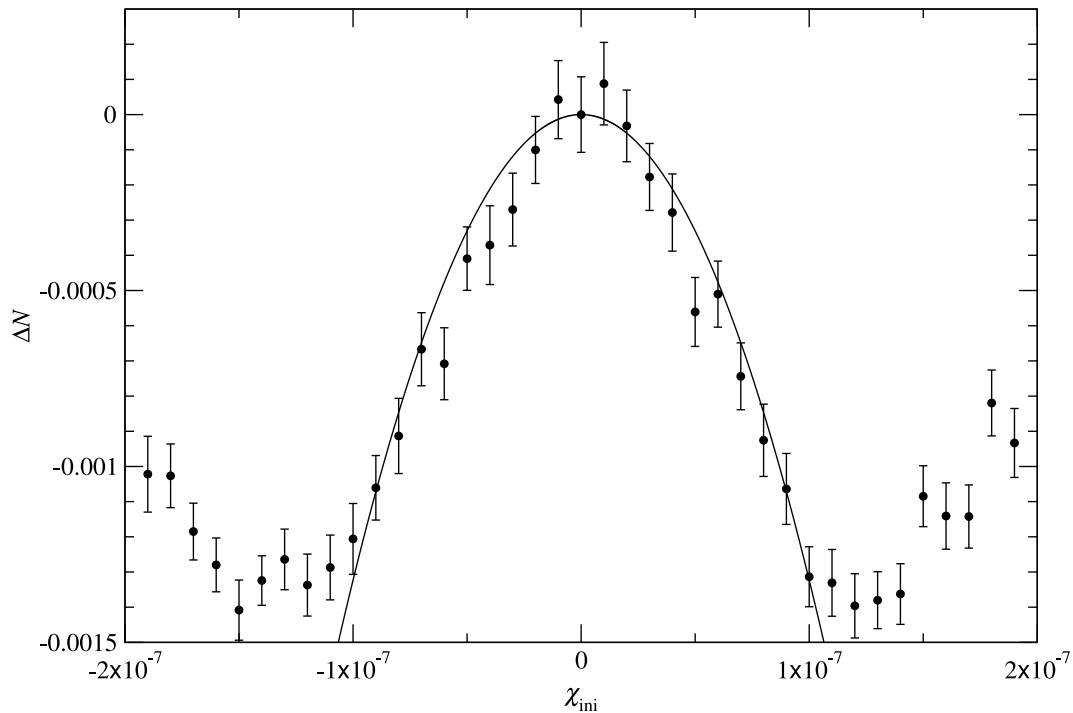}
 \caption{The dependence of $N$ on $\chi_{\rm ini}$ for $g^2/\lambda=1.875$ measured at $H=5.53\times 10^{-12}$. The curve shows a quadratic fit for low $N$ on $\chi_{\rm ini}$ to the function $N(\chi_{\rm ini})=N(0)+c\chi_{\rm ini}^2$. Due to the symmetry of the system simulations are only run for positive $\chi_{\rm ini}$ for all other $g^2/\lambda$.}
 \label{fig:mirror}
\end{figure}

From this data, we obtain the second derivative $\partial^2 N/\partial\chi_{\rm ini}^2$ by doing a fit at low $\chi_{\rm ini}$ with a quadratic function,
\begin{equation}
\label{equ:quadfit}
 N(\chi_{\rm ini})=N(0)+c\chi_{\rm ini}^2,
\end{equation}
so that $\partial^2N/\partial\chi_{\rm ini}^2=2c$. For each $\chi_{\rm ini}$ we repeated the simulation between 60 and 240 times, each with a different random realisation of the initial fluctuations. The averages for each $\chi_{\rm ini}$ can then be plotted (see for example figure \ref{fig:mirror}) and a best fit for the parameters $N(0)$ and $c$ in (\ref{equ:quadfit}) can be found. As the figure shows, the function (\ref{equ:quadfit}) is only a good fit for small $\chi_{\rm ini}$. We must therefore make a choice of which points to do the fit over. To do this, we start with a small number of points included in the fit and steadily add more points to the fit. When the statistical $\chi^2$ per degree of freedom grows significantly above one, points of that $\chi_{\rm ini}$ and above are not used.\footnote{There is overlapping notation here. `$\chi^2$ per degree of freedom' refers to the statistical $\chi^2$ technique and $\chi_{\rm ini}$ refers to the scalar field $\chi$.}

\begin{figure}
 \centering
 \includegraphics*[width=11cm]{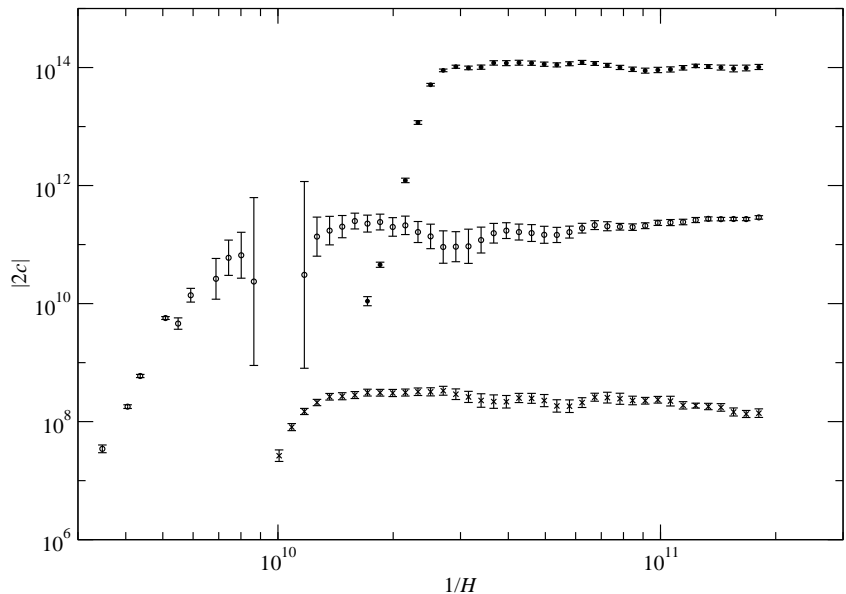}
 \caption{The evolution of the fit parameter $c$ in figure \ref{fig:mirror} and for two other choices of $g^2/\lambda$. Filled circles: $g^2/\lambda=2.7$. Unfilled circles: $g^2/\lambda=1.875$. Crosses: $g^2/\lambda=1.175$. Note that $c$ does not change at late times.}
 \label{fig:cgraph}
\end{figure}

In figure~\ref{fig:cgraph} we show how the result depends on the value of $H_*$ at which it is measured. The plot confirms that at late enough times the result is independent of $H_*$ as it should be.

\begin{table}
\caption{\label{tab:results}
The results from simulations are shown in the final column. The corresponding heights and positions in the resonance structure (figure \ref{fig:floquet1}) are shown for reference. $\mu_{\rm max}$ is the largest Floquet index for non-zero $\kappa^2$. Note that for many simulations this is at the largest scale in the lattice.}
\begin{indented}
\item[]\begin{tabular}{@{}lllll}
\br
$g^2/\lambda$ & $\mu_{\kappa=0}$ & $\mu_{\rm max}$ & $\kappa^2\left(\mu_{\rm max}\right) $ & $\frac{\partial^2N}{\partial\chi_{\rm ini}^2}$ \\
\mr
1.050 & 0.085 & 0.154 & 0.202 & $\lineup\m 10^{4.56\pm 0.07}$ \\ 
1.100 & 0.118 & 0.162 & 0.176 & $\lineup\m 10^{6.72\pm 0.08}$ \\ 
1.150 & 0.141 & 0.169 & 0.150 & $\lineup\m 10^{7.57\pm 0.05}$ \\ 
1.175 & 0.151 & 0.173 & 0.136 & $\lineup\m 10^{8.12\pm 0.05}$ \\ 
1.185 & 0.154 & 0.174 & 0.131 & $\lineup\m 10^{8.15\pm 0.05}$ \\ 
1.192 & 0.157 & 0.175 & 0.127 & $-10^{9.69\pm 0.12}$ \\ 
1.450 & 0.212 & 0.211 & 0.025 & $-10^{9.73\pm 0.08}$ \\ 
1.500 & 0.219 & 0.217 & 0.025 & $-10^{10.20\pm 0.06}$ \\ 
1.550 & 0.224 & 0.221 & 0.025 & $-10^{9.87\pm 0.15}$ \\ 
1.875 & 0.238 & 0.230 & 0.025 & $-10^{11.40\pm 0.02}$ \\ 
2.000 & 0.236 & 0.227 & 0.025 & $-10^{11.18\pm 0.15}$ \\ 
2.300 & 0.218 & 0.204 & 0.025 & $-10^{11.30\pm 0.09}$ \\ 
2.700 & 0.157 & 0.131 & 0.025 & $-10^{13.71\pm 0.10}$ \\
6.300 & 0.133 & 0.175 & 0.320 & $-10^{6.94\pm 0.20}$ \\
8.000 & 0.237 & 0.233 & 0.025 & $\lineup\m 10^{10.09 \pm 0.05}$ \\
9.500 & 0.151 & 0.135 & 0.025 & $-10^{11.30 \pm 0.03}$ \\
\br
\end{tabular}
\end{indented}
\end{table}

The final results measured at $H_*=5\times 10^{-12}M_{\rm Pl}$ are shown in table \ref{tab:results} and in figure~\ref{fig:redline}. Table~\ref{tab:results} shows that the sign of the result varies as suggested by the analytic result (\ref{eqn:analyicfin}). The figure also shows the analytic result in (\ref{eqn:analyicfin}), which is in very good agreement with our data, particularly in the first resonance band. This demonstrates that the calculation in Section~\ref{sec:ana} captures the relevant physics, and gives a potentially very useful way of calculating curvature perturbations in other models without having to carry out numerical simulations. Because of this, and because the analytic result covers all values of $g^2/\lambda$, we use it to draw some further conclusions.

In figure~\ref{fig:redline} we also show where the amplitude and the non-Gaussianity of the perturbations exceed the observed values according to (\ref{equ:cconstraint}) and (\ref{eqn:blfnltilt1}).
Comparing the data to the non-Gaussianity limits shows that when the large scale modes are within most of the first resonance band ($1<g^2/\lambda<3$) the prediction for $f_{\rm NL}$ is far outside observational bounds. Elsewhere the predicted $f_{\rm NL}$ is negligibly small. The ranges of $g^2/\lambda$ where $f_{\rm NL}$ is compatible with current data but potentially observable, i.e., $1\lesssim |f_{\rm NL}|\lesssim 100$, are extremely narrow: $1.0321<g^2/\lambda<1.0408$ and $2.9934<g^2/\lambda<2.9941$.

Within most of the first resonance band ($1.060\lesssim g^2/\lambda\lesssim 2.992$) the constraint (\ref{equ:cconstraint}) is not satisfied, meaning that the amplitude of the power spectrum due to the linear term in (\ref{eqn:zeta3}) also exceeds the observed value. This means that also the Gaussian perturbations are dominated by the contribution from preheating. By varying $\lambda$ their amplitude can be tuned to the observed level, and we plan to investigate this interesting possibility further in a future work. However, in this paper we simply conclude that for the usual choice of $\lambda$ these values of $g^2/\lambda$ are ruled out.

Figure~\ref{fig:redline} also shows that when the zero mode falls in the second resonance band, $6<g^2/\lambda<10$, and higher resonace bands, the constraint (\ref{equ:cconstraint}) is satisfied but preheating does not lead to observable non-Gaussianity.

As discussed in section~\ref{sec:ana}, our method has its limitations: the sharp spike close to $g^2/\lambda=3$, which is present both in the analytic and numerical results, is unphysical and can be interpreted as a finite-size effect. As shown in figures \ref{fig:floquet1} and \ref{fig:floquet2_log}, for these values of $g^2/\lambda$ the first peak in the Floquet index $\mu$ is narrow and includes only a few modes with low $\kappa$. It is these modes which make the dominant contribution to $\partial^2N/\partial\chi_{\rm ini}^2$, and as $g^2/\lambda$ approaches $3$ our finite simulation box contains fewer and fewer modes which are in resonance until the zero mode is artificially isolated by being the only mode falling in the first resonance band. This problem cannot be solved by using larger lattices, because if the lattice size $L$ exceeds the horizon size $1/H$, we cannot assume that the whole lattice is described by a homogeneous FRW metric and the results (\ref{eqn:blfnl}) and (\ref{eqn:blfnltilt1}) would not hold \cite{Lyth:2007jh}. The dashed red line in figure \ref{fig:redline} shows the prediction from (\ref{eqn:analyicfin}) for a box of infinite (and therefore unrealistic) size. The closeness of this to the prediction for a box of the size of our lattice simulations and to the results of the simulations shows that our conclusions do not depend heavily on the box size apart from in these sharp spikes. However, problems arising from the box size being slightly larger than the horizon could be addressed by introducing linear metric perturbations on the lattice as was done in \cite{BasteroGil:2007mm}, as long as deviations from homogeneity within the lattice are small.

\section{Conclusions}
In this paper we have presented full details of the method for calculating the curvature perturbations produced by non-equilibrium dynamics after the end of inflation which we introduced in \cite{Chambers:2007se}.
The method can be applied to many models, and we have demonstrated it by analysing the massless preheating model.

Our results show that preheating can have a very large effect on the curvature perturbations, but that it depends sensitively on the model and its parameters. In the case of massless preheating, the contribution is small if $g^2/\lambda$ is large, because the $\chi$ field is massive and its fluctuations are suppressed. For $g^2/\lambda\sim O(1)$ we find a non-trivial and interesting structure. If preheating is dominated by long-wavelength modes, it produces large non-Gaussian curvature perturbations which are incompatible with observational limits on $f_{\rm NL}$. For most of these values even the amplitude of the perturbations is too high. The amplitude can be reduced to an acceptable level by decreasing the value of $\lambda$, leading to a scenario in which the observed perturbations arise predominantly from preheating. We will investigate this interesting possibility in a future publication.

There are, however, narrow regions near $g^2/\lambda=1$ and $g^2/\lambda=3$ that are compatible with current observations~\cite{Komatsu:2008hk} but for which $f_{\rm NL}$ is large enough to be observed with future experiments such as the Planck satellite and can even saturate the current observational bounds.

It is important that preheating, a small addition to the simplest inflationary models which was originally introduced for very different reasons, can produce such levels of non-Gaussianity when slow roll inflation alone cannot. Unfortunately it also means that observation of non-Gaussianity can neither prove nor disprove these models.

What is also important is that even if $\chi$ is light, preheating produces no observable effects if it is dominated by inhomogeneous modes. These values of $g^2/\lambda$ are allowed in spite of the presence of isocurvature $\chi$ field fluctuations, because they are wiped out by preheating. 

It will be very interesting to see how well our findings generalize to other, more realistic inflationary models with preheating or other non-equilibrium phenomena. Our numerical method can be readily applied to any bosonic model, although this will be somewhat more costly since massless preheating has some special properties that make simulations particularly easy. Perhaps an easier way to study such models would be to use the analytic approximation we present in section~\ref{sec:ana}, which reproduces our numerical results and does not require numerical simulations.

\section*{Acknowledgments}

This work was supported by the Science and Technology Facilities Council and made use of the Imperial College High Performance Computing facilities. The authors would also like to thank Lev Kofman and Gary Felder for valuable discussions.

\appendix

\section{Power spectra at the beginning of simulations}
\label{app:spectra}

\subsection{Amplitude of perturbations}
The curvature perturbation generated during preheating is a function of the average value of the field $\chi$ over the simulation volume at the start of the simulation. The spectrum and other statistical properties of the curvature perturbation are therefore determined by the spectrum ${\cal P}_\chi$ of $\chi$ at the time when them simulation starts, which is shortly before the end of inflation.

During inflation the evolution of $\delta\chi_k$ is given by
\begin{equation}
\label{equ:modeeq}
 \ddot{\delta\chi_k}+3H\dot{\delta\chi_k}+g^2\phi^2\delta\chi_k=0.
\end{equation}
If we approximate the field to be massless, $g^2\phi^2\ll \left( 3H/2\right)^2$ then the modes are frozen once they leave the horizon and we have the standard result for a massless field \cite{liddlelyth},
\begin{equation}
\label{flatspectrum}
 {\cal P}_\chi(k)\equiv \frac{k^3}{2\pi^2}
 \left\vert\delta\chi_k\right\vert^2=\left. \frac{H^2}{4\pi^2}\right\vert_{k=aH}.
\end{equation}

\begin{equation}
\label{hwithn}
 H^2\simeq\frac{16}{3}\lambda M_{\rm Pl}^2N^2
\end{equation}
and
\begin{equation}
\label{phiwithn}
 \phi^2\simeq 8M_{\rm Pl}^2N.
\end{equation}
In these equations, $N$ measures the number of $e$-foldings before the end of inflation, but it should be noted that inflation actually ends at $N=3/2$ where the slow roll conditions fail as $\eta=1$.
The slow decrease of $H$ in (\ref{hwithn}) makes the spectrum (\ref{flatspectrum}) scale dependent.

The $\chi$ field is also not exactly massless during inflation, which introduces another source of scale dependence to ${\cal P}_\chi$. The Hubble damping term $3H$ in (\ref{equ:modeeq}) decreases more rapidly during inflation than the mass $g\phi$. At $N=N_{\rm crit}\equiv (2/3)g^2/\lambda$, when $g\phi=3H/2$, $\chi$ becomes underdamped.
A mode leaving the horizon before this will experience $N_k-N_{\rm crit}$ $e$-foldings of overdamped freeze-out followed by $N_{\rm crit}$ $e$-foldings of underdamped oscillations, where $N_k$ is the number of $e$-foldings before the end of inflation when the mode leaves the horizon \cite{Zibin:2000uw}. Modes leaving the horizon during the short underdamped period can be ignored, since they are never amplified by inflation. 

Our simulation starts at $N=N_{\rm sim}$. If $N_{\rm sim}>N_{\rm crit}$, which corresponds to small $g^2/\lambda$, then the entire underdamped period is calculated by our lattice simulations. In this case, the overdamped period adds a factor of \cite{Zibin:2000uw}
\begin{equation}
 \left\vert \exp \left[ -\int_{t_k}^{t_{\rm end}}\left( \frac{3H}{2}-\sqrt{\frac{9H^2}{4}-g^2\phi^2}\right)dt\right]\right\vert^2
\end{equation}
to (\ref{flatspectrum}). Substituting, (\ref{hwithn}) and (\ref{phiwithn}), $\dot\phi\simeq -\lambda\phi^3/3H$ and $dt=d\phi/\dot\phi$ leads to,
\begin{equation}
 \left\vert \exp \left[ -\int_{N_k}^{N_{\rm sim}}\left( \sqrt{\frac{9}{4}-\frac{3}{2}\frac{g^2}{\lambda}\frac{1}{N}}-\frac{3}{2}\right)dN\right]\right\vert^2.
\end{equation}
Solving the integral, (\ref{flatspectrum}) becomes \cite{Zibin:2000uw},
\begin{equation}
\label{eqn:overonly}
 {\cal P}_\chi(k)=\left. \frac{H^2}{4\pi^2}\right\vert_{k=aH}e^{-3F(N_k,N_{\rm sim})},
\end{equation}
where
\begin{eqnarray}
\fl F(N_k,N_{\rm sim})=N_k-N_{\rm sim}
+\sqrt{N_{\rm sim}}\sqrt{N_{\rm sim}-N_{\rm crit}}
-\sqrt{N_k}\sqrt{N_k-N_{\rm crit}}\nonumber\\
+N_{\rm crit}\log\left(\frac{\sqrt{N_k}+\sqrt{N_k-N_{\rm crit}}}
{\sqrt{N_{\rm sim}}+\sqrt{N_{\rm sim}-N_{\rm crit}}}\right).
\end{eqnarray}

In the case where $N_{\rm sim}<N_{\rm crit}$, i.e., for large $g^2/\lambda$, there is a similar overdamped contribution from $N_k$ to $N_{\rm crit}$ and also a contribution from the underdamped period between $N_{\rm crit}$ and $N_{\rm sim}$ \cite{Zibin:2000uw},
\begin{equation}
\label{eqn:underint}
 \left\vert \exp \left[ -\int_{t_{\rm crit}}^{t_{\rm end}} \frac{3H}{2}dt\right]\right\vert^2=\left\vert \exp \left[ -\frac{3}{2}\int_{N_{\rm crit}}^{N_{\rm sim}}dN\right]\right\vert^2= e^{-2\frac{g^2}{\lambda}+3N_{\rm sim}}
\end{equation}
giving
\begin{equation}
\label{withunder}
 {\cal P}_\chi(k)=\left. \frac{H^2}{4\pi^2}\right\vert_{k=aH}e^{-3F(N_k,N_{\rm crit})-2\frac{g^2}{\lambda}+3N_{\rm sim}}.
\end{equation}

As discussed in Section~\ref{sec:sepuni}, our assumption that the curvature perturbations are dominated by the inflaton field leads to the constraint (\ref{equ:cconstraint}) on the typical values of $\overline{\chi}_{\rm ini}$ and $\Delta\chi_{\rm ini}$. These are given by (\ref{equ:varchibar}) and (\ref{equ:vardelchi}). Changing the integration variable from $k$ to $N_k$, we can write the equations as
\begin{equation}
\label{equ:varchibarN}
 \left\langle \overline{\chi_{\rm ini}}^2\right\rangle =\int_{N_0}^{N_{\rm tot}}\mathcal P_\chi\left(k\right)
 \left(1-\frac{1}{N_k}\right)dN_k,
\end{equation}
and
\begin{equation}
\label{equ:vardelchiN}
 \left\langle \Delta{\chi_{\rm ini}}^2\right\rangle =\int_{N_{\rm sim}}^{N_0}\mathcal P_\chi\left(k\right)
 \left(1-\frac{1}{N_k}\right)dN_k,
\end{equation}
where $N$ is the number of $e$-foldings before the end of inflation, $N_0\approx 60$ is when the largest currently observable scales left the horizon, and the cutoff $N_{\rm tot}>N_0$ is the total number of $e$-foldings of inflation. Evaluating these integrals numerically, we find the constraint shown in figure \ref{fig:redline}.

At small $g^2/\lambda$ the integral diverges as $N_{\rm tot}\rightarrow \infty$, and therefore the constraint is depends on the total number of $e$-foldings $N_{\rm tot}$.
In the limit $g^2/\lambda\rightarrow 0$, we have $N_{\rm crit}=0$ and $F(N_k,N_{\rm sim})=0$, so that
\begin{equation}
{\cal P}_\chi(k)
=\frac{H^2}{4\pi^2}\approx \frac{4}{3\pi^2}\lambda M_{\rm Pl}^2N_k^2.
\end{equation}
We can therefore solve the integrals (\ref{equ:varchibarN}) and (\ref{equ:vardelchiN}) easily and find
\begin{eqnarray}
 \left\langle \overline{\chi_{\rm ini}}^2\right\rangle &\approx &
\frac{4}{9\pi^2}\lambda M_{\rm Pl}^2 N_{\rm tot}^3,\nonumber\\
  \left\langle \Delta{\chi_{\rm ini}}^2\right\rangle &\approx &
\frac{4}{9\pi^2}\lambda M_{\rm Pl}^2 N_{\rm 0}^3.
\end{eqnarray}
Substituting these into (\ref{equ:cconstraint}) leads to (\ref{equ:smallglimit}).

\subsection{Non-Gaussianity}

When making estimations of $f_{\rm NL}$ from (\ref{equ:fNLcalc}) we also need to know the power spectrum $\mathcal P_{\phi}$ of the inflaton $\phi$, which we assume to make the dominant contribution to
the curvature perturbation. Perturbations of $\phi$ evolve according to
\begin{equation}
\label{equ:modeeqphi}
 \ddot{\delta\phi_k}+3H\dot{\delta\phi_k}+3\lambda\phi^2\delta\phi_k=0,
\end{equation}
which is identical to (\ref{equ:modeeq}) if we replace $g^2\rightarrow 3\lambda$. This corresponds to $N_{\rm crit}=2$, and therefore the power spectrum is given by the analog of (\ref{eqn:overonly}),
\begin{equation}
\label{eqn:overonlyphi}
 {\cal P}_\phi(k)=\left. \frac{H^2}{4\pi^2}\right\vert_{k=aH}e^{\left.-3F(N_k,N_{\rm sim})
 \right|_{N_{\rm crit}=2}}.
\end{equation}

In order to estimate $f_{\rm NL}$ we must modify (\ref{equ:fNLcalc}) which was derived using the method described in \cite{Boubekeur:2005fj} for the case of non-scale-invariant power spectra. We split this into two parts. Firstly the we find the form of the ratio of power spectra ${\cal P}_\chi^3/{\cal P}_\zeta^2$ and then secondly we modify logarithmic factor.

For the case in which $N_{\rm sim}>N_{\rm crit}$ (the simulation begins before $\chi$ field enters the underdamped period) we substitute (\ref{eqn:overonly}) and (\ref{eqn:overonlyphi}) in to ${\cal P}_\chi^3/{\cal P}_\zeta^2$ using the slow roll solutions to estimate the scale-invariant component,
\begin{equation}
 \frac{{\cal P}_\chi^3}{{\cal P}_\zeta^2}\approx-\frac{16}{3\pi^2}\lambda
M_{\rm Pl}^6 e^{-9F(N_k,N_{\rm sim})+\left.6F(N_k,N_{\rm sim})
 \right|_{N_{\rm crit}=2}}.
\end{equation}
Making the approximations $N_k\gg N_{\rm crit}$ and $N_k\gg N_{\rm sim}$ the function $F(N_k,N_{\rm sim})$ becomes
\begin{equation}
 F(N_k,N_{\rm sim})=N_{\rm crit}\log\left(\sqrt{\frac{N_k}{N_{\rm sim}}}\right),
\end{equation} 
and therefore,
\begin{equation}
\label{eqn:ratio1}
\frac{{\cal P}_\chi^3}{{\cal P}_\zeta^2}\approx-\frac{16}{3\pi^2}\lambda
M_{\rm Pl}^6\left(\frac{N_k}{N_{\rm sim}}\right)^{3(2-g^2/\lambda)}.
\end{equation}

In the case in which $N_{\rm sim}<N_{\rm crit}$ (the simulation begins after the $\chi$ field has entered the underdamped period) we substitute (\ref{withunder}) and (\ref{eqn:overonlyphi}) in to ${\cal P}_\chi^3/{\cal P}_\zeta^2$:
\begin{equation}
 \frac{{\cal P}_\chi^3}{{\cal P}_\zeta^2}\approx-\frac{16}{3\pi^2}\lambda
M_{\rm Pl}^6 e^{-9F(N_k,N_{\rm crit})-6\frac{g^2}{\lambda}+9N_{\rm sim}+\left.6F(N_k,N_{\rm sim}) \right|_{N_{\rm crit}=2}},
\end{equation}
which in the large $N_k$ limit is
\begin{equation}
\label{eqn:ratio2}
 \frac{{\cal P}_\chi^3}{{\cal P}_\zeta^2}\approx-\frac{16}{3\pi^2}\lambda
M_{\rm Pl}^6
\left(\frac{6e^2N_k}{g^2/\lambda}\right)^{-3g^2/\lambda}
\left(\frac{N_k}{N_{\rm sim}}\right)^6 e^{9N_{\rm sim}}.
\end{equation} 

The origin of the logarithm in (\ref{equ:fNLcalc}) is in the estimation of the integral \cite{Boubekeur:2005fj}
\begin{equation}
\label{eqn:blint}
 \int^k_{a_0H_0} d^3q \frac{1}{q^3}\frac{1}{\left\vert q-k_1\right\vert^3}\frac{1}{\left\vert q+k_2\right\vert^3}{\cal P}_\chi (q){\cal P}_\chi(q-k_1){\cal P}_\chi(q+k_2).
\end{equation} 
We approximate $k_1\sim k_2\sim k$. For large $q$ the integral goes as $\sim \frac{1}{q^9}$, so we assume that the dominant contribution come from $q\ll k$ and the integral becomes
\begin{equation}
 \frac{{\cal P}_\chi(k)^2}{k^6}\int^k_{a_0H_0}d^3q\frac{1}{q^3}{\cal P}_\chi (q),
\end{equation} 
which in terms of $N$ is
\begin{equation}
 \frac{{\cal P}_\chi(k)^2}{k^6}\int_{N_k}^{N_0}dN_q{\cal P}_\chi(N_q)\left(1-\frac{1}{N_q}\right).
\end{equation}
At large $N_q$ we can drop the $\left(1-\frac{1}{N_q}\right)$ term and the power spectrum goes as
\begin{equation}
 {\cal P}_\chi\sim  \int^{N_0}_{N_k}dN_qN^2_q\left(\frac{N_q}{N_{\rm sim}}\right)^{-\frac{3}{2}N_{\rm crit}}\sim \int^{N_0}_{N_k}dN_q N_q^{2-\frac{g^2}{\lambda}}.
\end{equation} 
From this it can be seen that as found previously \cite{Zibin:2000uw} the spectrum is only scale invariant for $g^2/\lambda\approx 2$ in which case (\ref{eqn:blint}) is proportional to
\begin{equation}
 \int^{N_0}_{N_k}dN_k\sim N_0-N_k \sim \ln\left(\frac{k}{a_0H_0}\right).
\end{equation} 
In general we have,
\begin{equation}
\fl \int^{N_0}_{N_k}dN_q N_q^{2-\frac{g^2}{\lambda}}\sim N_0^{2-\frac{g^2}{\lambda}}-N_k^{2-\frac{g^2}{\lambda}}\nonumber
\sim \left(\log\frac{k}{\sqrt{\lambda}M_{\rm Pl}}\right)^{2-\frac{g^2}{\lambda}}-\left(\log\frac{a_0H_0}{\sqrt{\lambda}M_{\rm Pl}}\right)^{2-\frac{g^2}{\lambda}}.
\end{equation}

Combining this with (\ref{equ:fNLcalc}) and in the $N_{\rm sim}>N_{\rm crit}$ case with (\ref{eqn:ratio1})  we have
\begin{eqnarray}
\fl f_{\rm NL}\approx -\frac{5}{9\pi^2}\left(\frac{\partial^2N}{\partial\chi_{\rm ini}^2}\right)^3\lambda
M_{\rm Pl}^6\left(\frac{N_k}{N_{\rm sim}}\right)^{3(2-g^2/\lambda)}\nonumber\\
\times\left(\left(\log\frac{k}{\sqrt{\lambda}M_{\rm Pl}}\right)^{2-\frac{g^2}{\lambda}}-\left(\log\frac{a_0H_0}{\sqrt{\lambda}M_{\rm Pl}}\right)^{2-\frac{g^2}{\lambda}}\right).
\end{eqnarray} 
In the $N_{\rm sim}<N_{\rm crit}$ case we substitute (\ref{eqn:ratio2}) into (\ref{equ:fNLcalc}):
\begin{eqnarray}
\fl f_{\rm NL}\approx -\frac{5}{9\pi^2}\left(\frac{\partial^2N}{\partial\chi_{\rm ini}^2}\right)^3\lambda
M_{\rm Pl}^6\left(\frac{6e^2N_k}{g^2/\lambda}\right)^{-3g^2/\lambda}
\left(\frac{N_k}{N_{\rm sim}}\right)^6 e^{9N_{\rm sim}}\nonumber\\
\times\left(\left(\log\frac{k}{\sqrt{\lambda}M_{\rm Pl}}\right)^{2-\frac{g^2}{\lambda}}-\left(\log\frac{a_0H_0}{\sqrt{\lambda}M_{\rm Pl}}\right)^{2-\frac{g^2}{\lambda}}\right).
\end{eqnarray}
Dropping the numerical constant and the logarithms leads to (\ref{eqn:blfnltilt1}).

\section{Numerical algorithm}
\label{app:rk}

Here we will present the numerical algorithm used in our lattice simulation to evolve the field equations and the Friedmann equations. This is done by the numerical integration of (\ref{phi_eq_motion}) and (\ref{equ:inhomo3}). The variables to be solved for are $\phi^{ijk}(\tau)$, $\chi^{ijk}(\tau)$ and $a(\tau)$, for all $i$, $j$ and $k$ where $i\in \left\lbrace 0,...,S-1\right\rbrace$, $j\in \left\lbrace 0,...,S-1\right\rbrace$ and $k\in \left\lbrace 0,...,S-1\right\rbrace$ indicate the position in the lattice, and there are $S^3$ points in the lattice. 
We use conformal time $\tau$ defined by
$d\tau=dt/a(t)$, and $'$ will indicate the derivative with respect to $\tau$.
The conformal time step is denoted by $\delta\tau$ and the comoving lattice spacing by $\delta x$.

The field and Friedmann equations are coupled in a leapfrog-like fashion, by defining the scale factor $a$ at half-way between time steps of the field evolution.
To achieve this, the scale factor is first evolved by half a timestep using slow roll. Making the usual assumptions, $a'(\tau_{\rm ini})=\sqrt{\frac{\lambda\phi^4_{\rm ini}}{12 M_{\rm Pl}}}a(\tau_{\rm ini})^2$, and then,
\begin{equation}
 a\left(\tau_{\rm ini}+\frac{\delta\tau}{2}\right) =a(\tau_{\rm ini})+\frac{\delta\tau}{2}a'(\tau_{\rm ini}).
\end{equation}
Slow roll is not imposed after this initial half-step. The fields are then evolved one timestep to be half a timestep ahead of $a(\tau)$, and then $a(\tau)$ is evolved one timestep to be half a timestep ahead of the fields, and so on.

The field evolution is by a standard fourth order Runge-Kutta method \cite{abramsteg}. The first derivatives are given the status of independent variables: $\phi^{ijk}_{\rm p}=\left(\phi^{ijk}\right)'$ and $\chi^{ijk}_{\rm p}=\left(\chi^{ijk}\right)'$. Therefore at the $n$th timestep:
\begin{eqnarray}
\nonumber 
\fl \left(\phi^{ijk}\right)''_n=\left(\phi^{ijk}_{{\rm p}}\right)'_n=f_\phi\left(\phi^{ijk}_{\ \ n},\phi^{ijk}_{{\rm p}\ n},\chi^{ijk}_{\rm p},\chi^{ijk}_{{\rm p}\ n}\right)\\
=\nabla^2\phi^{ijk}_{\ \ n}- 2\frac{a'_{n+\frac{1}{2}}}{a_{n+\frac{1}{2}}}\phi^{ijk}_{{\rm p}\ n}- a_{n+\frac{1}{2}}^2\left(\lambda \left(\phi^{ijk}_{\ \ n}\right)^2+g^2 \left(\chi^{ijk}_{\ \ n}\right)^2\right)\phi^{ijk}_{\ \ n},\\
\nonumber 
\fl \left(\chi^{ijk}\right)''_n=\left(\chi^{ijk}_{{\rm p}}\right)'_n=f_\chi\left(\phi^{ijk}_{\ \ n},\phi^{ijk}_{{\rm p}\ n},\chi^{ijk}_{\ \ n},\chi^{ijk}_{{\rm p}\ n}\right)\\
=\nabla^2\chi^{ijk}_{\ \ n}- 2\frac{a'_{n+\frac{1}{2}}}{a_{n+\frac{1}{2}}}\chi^{ijk}_{{\rm p}\ n}- a_{n+\frac{1}{2}}^2 g^2 \left(\phi^{ijk}_{\ \ n}\right)^2{\chi^{ijk}_{\ \ n}},
\end{eqnarray}
where for $X\in \{\phi,\chi\} $,
\begin{eqnarray}
\fl \nabla^2X^{ijk}_{\ \ n}=\frac{1}{\delta x^2}\left( X^{[i+1]jk}_{\ \ n}+X^{[i-1]jk}_{\ \ n}+X^{i[j+1]k}_{\ \ n}\right.\nonumber\\
\left. +X^{i[j-1]k}_{\ \ n}+X^{ij[k+1]}_{\ \ n}+X^{ij[k-1]}_{\ \ n}-6X^{ijk}_{\ \ n}\right).
\end{eqnarray}
The square brackets indicate addition or subtraction in modulus $S$. We the define the Runge-Kutta parameters:
\begin{eqnarray}
\nonumber R^{ijk}_{11}&=&\phi^{ijk}_{{\rm p}\ n}\\
\nonumber R^{ijk}_{12}&=&f_\phi\left(\phi^{ijk}_{\ \ n},\phi^{ijk}_{{\rm p}\ n},\chi^{ijk}_{\rm p},\chi^{ijk}_{{\rm p}\ n}\right)\\
\nonumber R^{ijk}_{13}&=&\chi^{ijk}_{{\rm p}\ n}\\
\nonumber R^{ijk}_{14}&=&f_\chi\left(\phi^{ijk}_{\ \ n},\phi^{ijk}_{{\rm p}\ n},\chi^{ijk}_{\rm p},\chi^{ijk}_{{\rm p}\ n}\right)\\
\nonumber R^{ijk}_{21}&=&\phi^{ijk}_{{\rm p}\ n}+\frac{1}{2}R^{ijk}_{12}\\
\nonumber R^{ijk}_{22}&=&f_\phi\left(\phi^{ijk}_{\ \ n}+\frac{1}{2}R^{ijk}_{11},\phi^{ijk}_{{\rm p}\ n}+\frac{1}{2}R^{ijk}_{12},\chi^{ijk}_{\rm p}+\frac{1}{2}R^{ijk}_{13},\chi^{ijk}_{{\rm p}\ n}+\frac{1}{2}R^{ijk}_{14}\right)\\
\nonumber R^{ijk}_{23}&=&\chi^{ijk}_{{\rm p}\ n}+\frac{1}{2}R^{ijk}_{14}\\
\nonumber R^{ijk}_{24}&=&f_\chi\left(\phi^{ijk}_{\ \ n}+\frac{1}{2}R^{ijk}_{11},\phi^{ijk}_{{\rm p}\ n}+\frac{1}{2}R^{ijk}_{12},\chi^{ijk}_{\rm p}+\frac{1}{2}R^{ijk}_{13},\chi^{ijk}_{{\rm p}\ n}+\frac{1}{2}R^{ijk}_{14}\right)\\
\nonumber R^{ijk}_{31}&=&\phi^{ijk}_{{\rm p}\ n}+\frac{1}{2}R^{ijk}_{22}\\
\nonumber R^{ijk}_{32}&=&f_\phi\left(\phi^{ijk}_{\ \ n}+\frac{1}{2}R^{ijk}_{21},\phi^{ijk}_{{\rm p}\ n}+\frac{1}{2}R^{ijk}_{22},\chi^{ijk}_{\rm p}+\frac{1}{2}R^{ijk}_{23},\chi^{ijk}_{{\rm p}\ n}+\frac{1}{2}R^{ijk}_{24}\right)\\
\nonumber R^{ijk}_{33}&=&\chi^{ijk}_{{\rm p}\ n}+\frac{1}{2}R^{ijk}_{24}\\
\nonumber R^{ijk}_{34}&=&f_\chi\left(\phi^{ijk}_{\ \ n}+\frac{1}{2}R^{ijk}_{21},\phi^{ijk}_{{\rm p}\ n}+\frac{1}{2}R^{ijk}_{22},\chi^{ijk}_{\rm p}+\frac{1}{2}R^{ijk}_{23},\chi^{ijk}_{{\rm p}\ n}+\frac{1}{2}R^{ijk}_{24}\right)\\
\nonumber R^{ijk}_{41}&=&\phi^{ijk}_{{\rm p}\ n}+R^{ijk}_{32}\\
\nonumber R^{ijk}_{42}&=&f_\phi\left(\phi^{ijk}_{\ \ n}+R^{ijk}_{31},\phi^{ijk}_{{\rm p}\ n}+R^{ijk}_{32},\chi^{ijk}_{\rm p}+R^{ijk}_{33},\chi^{ijk}_{{\rm p}\ n}+R^{ijk}_{34}\right)\\
\nonumber R^{ijk}_{43}&=&\chi^{ijk}_{{\rm p}\ n}+R^{ijk}_{34}\\
R^{ijk}_{44}&=&f_\chi\left(\phi^{ijk}_{\ \ n}+R^{ijk}_{31},\phi^{ijk}_{{\rm p}\ n}+R^{ijk}_{32},\chi^{ijk}_{\rm p}+R^{ijk}_{33},\chi^{ijk}_{{\rm p}\ n}+R^{ijk}_{34}\right).
\end{eqnarray}
From these we can find,
\begin{eqnarray}
\nonumber \phi^{ijk}_{\ \ n+1}&=&\phi^{ijk}_{\ \ n}+\frac{\delta\tau}{6}\left( R^{ijk}_{11}+2R^{ijk}_{21}+2R^{ijk}_{31}+R^{ijk}_{41}\right)\\
\nonumber \phi^{ijk}_{{\rm p}\ n+1}&=&\phi^{ijk}_{{\rm p}\ n}+\frac{\delta\tau}{6}\left( R^{ijk}_{12}+2R^{ijk}_{22}+2R^{ijk}_{32}+R^{ijk}_{42}\right)\\
\nonumber \chi^{ijk}_{\ \ n+1}&=&\chi^{ijk}_{\ \ n}+\frac{\delta\tau}{6}\left( R^{ijk}_{13}+2R^{ijk}_{23}+2R^{ijk}_{33}+R^{ijk}_{43}\right)\\
\chi^{ijk}_{{\rm p}\ n+1}&=&\chi^{ijk}_{{\rm p}\ n}+\frac{\delta\tau}{6}\left( R^{ijk}_{14}+2R^{ijk}_{24}+2R^{ijk}_{34}+R^{ijk}_{44}\right).
\end{eqnarray}
The evolution of $a$ is by an Euler method. At each timestep we have the sum,
\begin{eqnarray}
\nonumber \fl \rho_n=\sum^{S-1}_{i=0}\sum^{S-1}_{j=0}\sum^{S-1}_{k=0}\left(  \frac{1}{2a_{n-\frac{1}{2}}^2}\left(\phi^{ijk}_{{\rm p}\ n}\right)^2+\frac{1}{2a_{n-\frac{1}{2}}^2}\left(\chi^{ijk}_{{\rm p}\ n}\right)^2 +\frac{1}{4}\lambda\left(\phi^{ijk}_{\ \ n}\right)^4+\frac{1}{2}g^2\left(\phi^{ijk}_{\ \ n}\right)^2\left(\chi^{ijk}_{\ \ n}\right)^2\right.\\
\left. +\frac{1}{2a_{n-\frac{1}{2}}^2}\left( \nabla {\phi^{ijk}_{\ \ n}}\right)^2+\frac{1}{2a_{n-\frac{1}{2}}^2}\left(\nabla {\chi^{ijk}_{\ \ n}}\right)^2\right),
\end{eqnarray}
where for $X\in \{\phi,\chi\} $,
\begin{equation}
\fl \nabla X^{ijk}_{\ \ n}=\frac{1}{2 \delta x}\left( X^{[i+1]jk}_{\ \ n}+X^{[j+1]jk}_{\ \ n}+X^{i[k+1]k}_{\ \ n}+X^{i[i-1]k}_{\ \ n}+X^{ij[j-1]}_{\ \ n}+X^{ij[k-1]}_{\ \ n}\right).
\end{equation}
Putting this into the Friedmann equation gives,
\begin{equation}
 a_{n+\frac{1}{2}}=a_{n-\frac{1}{2}}+\delta\tau\sqrt{\frac{\rho_n}{3M_{\rm Pl}^2}}a_{n-\frac{1}{2}}^2.
\end{equation}

\section*{References}
\bibliography{paper}

\end{document}